\documentstyle[mncite,psfig]{mn}

\title[On the nature of SW Sex]{On the nature of SW Sex}

\author[V.\,S.\ Dhillon et al.]{V.\,S.\ Dhillon$^{1}$, 
T.\,R.\ Marsh$^{2}$ and D.\,H.\,P.\ Jones$^{1}$\\
$^1$Royal Greenwich Observatory, Madingley Road, Cambridge CB3 0EZ 
(vsd@ast.cam.ac.uk, dhpj@ast.cam.ac.uk) \\
$^2$University of Southampton, Department of Physics and Astronomy, Highfield, 
Southampton SO17 1BJ (trm@phastr.soton.ac.uk)}

\date{Accepted 1997 June 19. Received 1997 April 25; in original form
1997 February 3}

\begin{document}
\maketitle

\begin{abstract} 
We present spectrophotometry of the eclipsing nova-like variable SW Sex. 
The continuum is deeply eclipsed and shows asymmetries due to the
presence of a bright spot. We derive a new ephemeris and, by measuring
the eclipse width, we are able to constrain the inclination to 
$i>75^\circ$ and the disc radius to $R_D>0.6L_1$. In common with other
members of its class (of which it is the proto-type), SW Sex shows
single-peaked emission lines which show transient absorption
features and large phase shifts in their radial velocity curves. In
addition, the light curves of the emission lines show a reduction
in flux around phase 0.5 and asymmetric eclipse profiles which are not as
deep as the continuum eclipse. Using Doppler tomography, we find that 
most of the line emission in SW Sex appears to originate from three 
sources: the secondary star, the accretion disc and an extended bright
spot. The detection of the red star allows us to constrain the 
radial-velocity semi-amplitude of the secondary to
$K_R>180$~km\,s$^{-1}$ and hence the component masses to 
$M_1\sim 0.3-0.7 M_{\odot}$ and $M_2<0.3 M_{\odot}$. 

The Doppler maps suggest a simple new
model for SW Sex in which the dominance of single-peaked 
line emission from the bright spot over the weak double-peaked disc 
emission gives SW Sex its single-peaked profiles and forces the 
radial-velocity curves to follow the motion of the bright spot and thus 
exhibit large phase shifts. The transient absorption features in the 
Balmer-line profiles are mostly artifacts of the complex intertwining of the
emission components from the secondary star, bright spot and accretion
disc and involve little true absorption. While the accretion disc and
secondary star components of this model appear to be secure, the
dominant bright-spot component fails in one important area -- its 
inconsistency with the Balmer-line light curves. The eclipse profile 
requires the material emitting the Balmer lines to be eclipsed as early 
as phase 0.8, but which is 
not as deeply eclipsed as the continuum, exhibits a flat bottomed eclipse and 
then comes out of eclipse very sharply around phase 0.05. Although it
is possible to explain the early ingress with a raised disc rim 
downstream from the bright-spot, the rapid egress is difficult to 
account for without speculating that either there are regions of strong
Balmer absorption in the disc whose changing visibility during 
eclipse alters the shape of the light curve or that 
there is Balmer emission from above the orbital plane which
shares the velocity of the bright-spot. 
\end{abstract} 


\section{Introduction}
\label{sec:introduction}

Cataclysmic variables (CVs) are close binary stars consisting of a red dwarf
secondary transferring material on to a white dwarf primary via an accretion
disc or magnetic accretion stream. Nova-like variables are defined as CVs 
which have never been observed to undergo nova or dwarf-nova type outbursts
(see \pcite{warner95} and \pcite{dhillon96} for reviews). This is due to the 
fact that nova-likes are believed to harbour steady-state accretion discs, one
of the foundations of accretion disc theory \cite{pringle81}.
Recent advances in accretion disc research have been based on studying 
nova-like variables (e.g. \pcite{rutten93}), but it has become increasingly 
clear that there may be problems with this approach due to the SW Sex
phenomenon, which throws into question even the very existence of accretion 
discs in these systems \cite{williams89}. 

The SW Sex phenomenon, a term first coined by \scite{thorstensen91a}, is
peculiar to a group of high-inclination nova-like variables with periods in 
the range 3--4 hr. All of these so-called SW Sex stars exhibit 
strong, single-peaked Balmer, He\,{\small I} and He\,{\small II} emission 
lines which, with the 
exception of He\,{\small II}, remain largely unobscured during primary 
eclipse. This is in stark contrast to standard accretion disc theory, 
which predicts that emission lines from high-inclination discs should appear 
double-peaked and be eclipsed once every orbital period (e.g. \pcite{horne86}).
In addition, the emission lines in SW Sex stars show strong phase-dependent 
absorption features and their radial velocity curves exhibit significant phase
shifts relative to photometric minimum.
Three competing models claim to satisfy these observational constraints, with 
varying degrees of success: an accretion disc wind \cite{honeycutt86}, 
magnetic accretion \cite{williams89} or a bright-spot overflow 
(\pcite{hellier94}). 

With the intent of discriminating between these opposing models, we obtained 
spectrophotometry of the proto-type of the class, SW Sex. This object (also 
known as PG\,1012--029) was one of 22 CVs discovered by their UV excesses
in photographic films obtained for the Palomar-Green survey \cite{green86}.
Follow-up spectroscopic observations of this $V\sim14.8$~mag object by 
\scite{green82} revealed a high-excitation emission-line spectrum, suggesting 
a nova-like variable classification. The first extensive photometric and 
spectroscopic study of SW Sex was presented by \scite{penning84}, who 
discovered 1.9~mag eclipses recurring on a 3.24~hr orbital period. 
This work was followed by the higher-resolution spectroscopic 
studies of Honeycutt et al. (1986; see also 
\pcite{kaitchuck94}) and \scite{williams89}, who reached different 
conclusions as to the nature of 
these systems, the former invoking an accretion disc wind and the latter a 
magnetic accretion stream to explain the single-peaked, uneclipsed emission
line profiles. The phase dependent absorption features observed by
\scite{honeycutt86} were studied in greater detail by \scite{szkody90}, 
but they were unable to account for the behaviour with a simple model. 
High-speed photometry by \scite{rutten92b}
revealed a prominent bright spot in SW Sex (also seen by \pcite{ashoka94}) 
and the resulting eclipse maps 
showed a flatter run of effective temperature with radial distance from disc 
centre than that predicted by steady-state disc theory. Despite the more
recent efforts of \scite{still95}, who presented a high signal-to-noise but 
relatively low spectral-resolution study of SW Sex, no real progress has been 
made in unravelling the SW Sex phenomenon, a situation which we hope to redress
in the present paper. 

\section{Observations}
\label{sec:observations} 

We obtained a total of 83 spectra of SW Sex on the night of 1990 January 18
(orbital phases 26456.5686--26457.6282) and 75 spectra on 1990 January 
19 (orbital phases 26464.0394--26464.9781)
with the 2.5-m Isaac Newton Telescope (INT) on La Palma. The exposures 
were all 120-s long with about 9-s dead-time for the archiving of data. The 
Intermediate Dispersion Spectrograph (IDS) coupled with the Image Photon 
Counting System (IPCS, camera format 2048$\times$130 
pixel) and a grating of 1200 line\,mm$^{-1}$ gave a wavelength coverage of 
approximately 4025--5050\,\AA\ at 1.1\,\AA\ 
($\simeq$75 km\,s$^{-1}$) resolution. 
Comparison arc spectra were taken every $\sim$40\,min to calibrate the 
wavelength scale and instrumental flexure. We were able to correct for slit 
losses by placing a nearby comparison star on the 140$\times$1.1 arcsec$^{2}$ 
slit and obtaining a photometric spectrum through a wide (5.3~arcsec) slit. 
We also took a spectrum of the standard star PG\,0823+546 \cite{massey88}
to correct the large-scale instrumental response and place the data on
an absolute flux scale. 

\section{Data Reduction} 
\label{sec:datared} 

The first step in the data reduction was the division of each frame by a 
normalized tungsten-lamp flat-field, which we used to correct for medium-scale
sensitivity variations of the detector; the large-scale variations were removed
using the standard star, while the small-scale variations were removed by
the IPCS scanning coils \cite{jorden86}. The sky was removed by subtracting
third-order polynomial fits to sky regions on either side of the SW Sex 
spectra. Extracted spectra were then obtained by summing across 
the stellar profile. The same procedure was applied to the comparison
star spectra. The wavelength scale for each spectrum was interpolated from the 
wavelength scales of two neighbouring arc spectra, extracted from the same 
location on the detector as the object. The rms scatter of the sixth-order
polynomial fits to the arc lines was always $<$0.1\,\AA. We applied a standard
correction for extinction. Our final procedure was to correct for large-scale
instrumental sensitivity variations and slit losses in order to obtain 
absolute fluxes. Each spectrum was first divided by a spline fit
to the ratio of the observed count rate to the tabulated flux of PG\,0823+546.
The comparison star was too faint to allow for a wavelength-dependent 
slit-loss correction, so each SW Sex spectrum was then divided by the
total flux in the corresponding comparison star spectrum and multiplied by the
total flux in the wide slit comparison star spectrum.

\section{Results}
\label{sec:results}

\subsection{Ephemeris}
\label{sec:ephem}

\begin{table}
\caption{Eclipse observations of SW Sex. The differences between the observed
and calculated times of mid-eclipse are given by O--C. The first 6 eclipse 
timings have been reproduced from \protect\scite{penning84} and the last 6 from
\protect\scite{ashoka94}. The remaining 9 eclipse timings have been kindly 
provided by {\protect Ren\'{e}} Rutten (private communication).}
{\bf
\begin{center} 
\begin{tabular}{@{}crr}
 & & \\
\multicolumn{1}{c}{\bf HJD (mid-eclipse)} &
\multicolumn{1}{c}{\bf Cycle Number} &
\multicolumn{1}{c}{\bf O--C} \\
\multicolumn{1}{c}{\bf (2\,440\,000+)} &
\multicolumn{1}{c}{\boldmath ($E$)} &
\multicolumn{1}{c}{\bf (s)} \\
& & \\
4339.65087    &    0 \ \ \ \ \ \     & 4.3    \\
4340.73055    &    8 \ \ \ \ \ \     & 19.2   \\
4348.82649    &    68 \ \ \ \ \ \    & --12.4 \\
4631.92758    &    2166 \ \ \ \ \ \  & 11.7   \\
4676.86195    &    2499 \ \ \ \ \ \  & 0.9    \\
4721.79636    &    2832 \ \ \ \ \ \  & --6.4  \\
7566.56813    &    23914 \ \ \ \ \ \ & --11.8 \\
7615.41619    &    24276 \ \ \ \ \ \ & 18.6   \\
7615.55065    &    24277 \ \ \ \ \ \ & --22.8 \\
7616.49516    &    24284 \ \ \ \ \ \ & --27.9 \\
7618.51922    &    24299 \ \ \ \ \ \ & --29.3 \\
7619.46374    &    24306 \ \ \ \ \ \ & --33.5 \\
7620.40856    &    24313 \ \ \ \ \ \ & --11.8 \\
7621.48834    &    24321 \ \ \ \ \ \ & 11.7   \\
7622.43257    &    24328 \ \ \ \ \ \ & --17.6 \\
7921.32167    &    26543 \ \ \ \ \ \ & 25.1   \\
7921.45633    &    26544 \ \ \ \ \ \ & 1.0    \\
7950.19842    &    26757 \ \ \ \ \ \ & 18.8   \\
7950.33321    &    26758 \ \ \ \ \ \ & 6.0    \\
8306.30100    &    29396 \ \ \ \ \ \ & 25.9   \\
8306.43599    &    29397 \ \ \ \ \ \ & 30.3   \\
\end{tabular}
\end{center}
}
\label{tab:ephem}
\end{table}
We derived an ephemeris for SW Sex by adding the 9 times of mid-eclipse 
observed by Ren\'{e} Rutten (private communication; see Table~\ref{tab:ephem})
to the 6 eclipse timings of \scite{penning84} and the 6 of \scite{ashoka94}. A 
linear least-squares fit to these data yields the following ephemeris:
\begin{flushleft}
$
\begin{array}{rrrl}
T_{\rm mid-eclipse} = & \!\!\!\!\! {\em HJD}\,\,2\,444\,339.650821
& \!\! + \,\, 0.1349384229 & \!\!\!\!\! E \\
& \!\! \pm \,\, 0.000043 & \!\! \pm \,\, 0.0000000020 & \\
\end{array}
$
\end{flushleft}
The errors in the ephemeris have been derived using uncertainties of
0.0001 days in the times of mid-eclipse. Only the times of mid-eclipse 
provided by Ren\'{e} Rutten actually had published uncertainty values.
We justify the use of the same errors for the data of 
\scite{penning84} and \scite{ashoka94} because the former quoted times
of mid-eclipse to 4 decimal places (implying the error is in the last digit)
and the data of \scite{ashoka94} are qualitatively similar to the other
two datasets (implying the error must be similar). Note that we did not
include our two spectrophotometrically-measured eclipse timings as the 
uncertainties on these points are much greater than the 
photometrically-measured points given in Table~\ref{tab:ephem} 
(due to the poorer time resolution) and our points were also 
observed very close in time to those of \scite{ashoka94}. 

The differences between the observed and calculated times of mid-eclipse 
(O--C) are 
given in Table~\ref{tab:ephem}. Given that most of the points cluster around
two widely-separated epochs and that the typical O--C values derived from the
linear ephemeris are of the same order as the uncertainties in the 
mid-eclipse timings, we do not think a search for 
variations in the orbital period (e.g. by fitting a parabola, 
see \pcite{pringle75}) is justified. 

\subsection{Inclination and disc radius}
\label{sec:inclrad}

\begin{figure}
\centerline{\psfig{figure=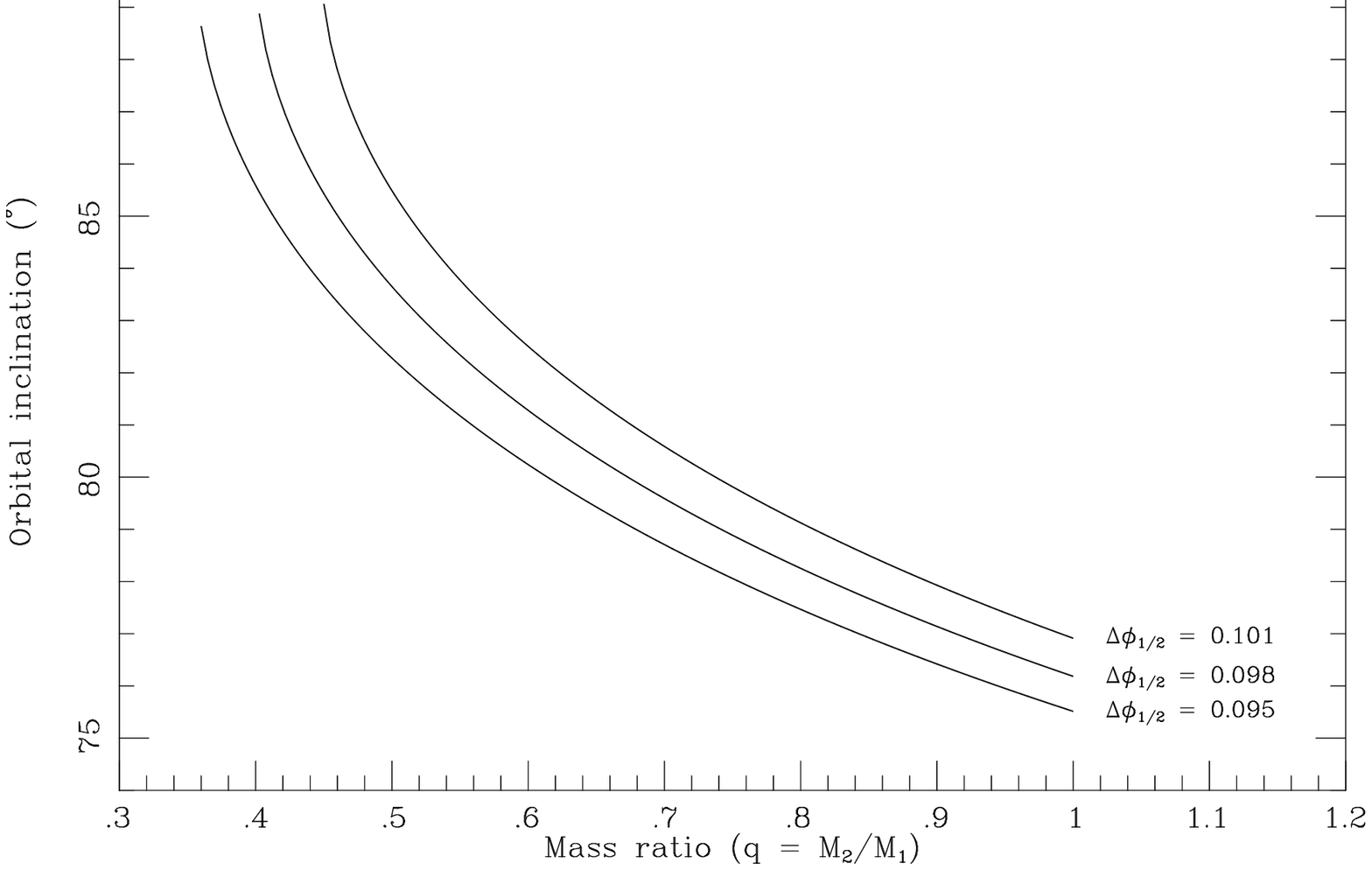,width=8.4cm,rheight=5.25cm}}
\centerline{\psfig{figure=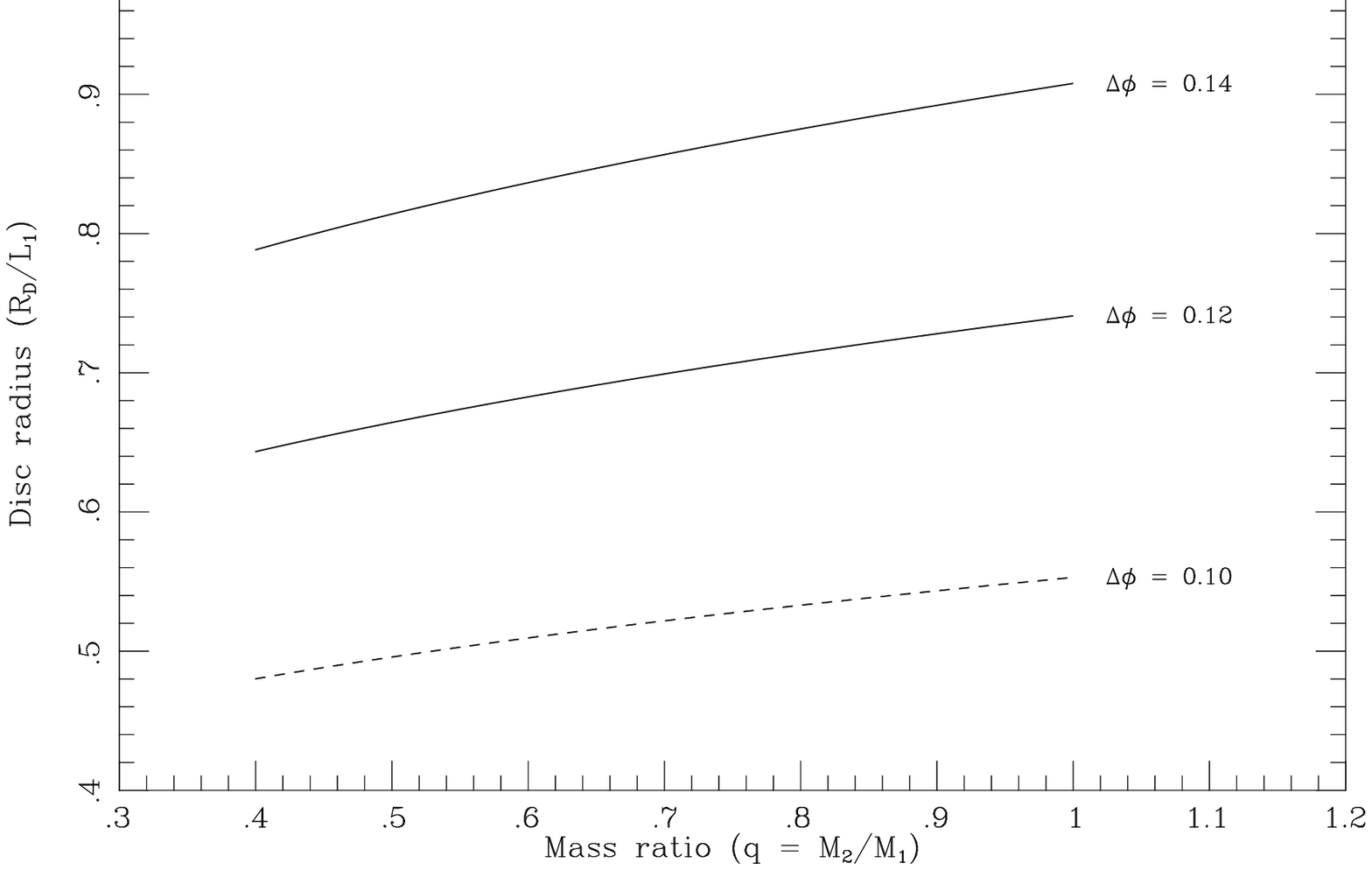,width=8.4cm,rheight=5.25cm}}
\caption{Upper panel -- orbital inclination as a function of mass ratio for
three different values of the eclipse width, $\Delta\phi_{1/2}=$\,0.095, 0.098
and 0.101. Lower panel -- accretion disc radius as a function of mass ratio 
for three different values of the eclipse width, $\Delta\phi=$\,0.10, 0.12
and 0.14.}
\label{fig:inclrad}
\end{figure}

By considering the geometry of a point eclipse by a spherical body, it is
possible to determine the inclination, $i$, of a binary system through the
relation
\begin{equation}
\left(\frac{R_2}{a}\right)^{2} = \sin^{2}(\pi\Delta\phi_{1/2}) + \cos^2(\pi\Delta\phi_{1/2})\cos^2{i},
\label{eqn:incq}
\end{equation}
where $R_2/a$ is defined as the volume radius of the secondary star and can
be expressed solely as a function of the mass ratio, $q=M_2/M_1$ 
\cite{eggleton83}:
\begin{equation}
\frac{R_2}{a} = \frac{0.49q^{2/3}}{0.6q^{2/3}+\ln(1+q^{1/3})}. 
\label{eqn:eggleton}
\end{equation}
$\Delta\phi_{1/2}$ is the mean phase full-width of eclipse at half the
out-of-eclipse intensity, which from the sum of the quasi-$B$-band
eclipses of \scite{rutten92b} was measured to be 
$\Delta\phi_{1/2}=0.098\pm0.003$ (Ren\'{e} Rutten, private communication). 
This figure is in good agreement with the eclipse widths derived from a 
visual inspection of Fig. 8 of \scite{penning84} and Fig. 8 of 
\scite{ashoka94}, which have a value of $\Delta\phi_{1/2}\sim0.1$. 
The upper panel of Fig.~\ref{fig:inclrad} shows 
the $q$ versus $i$ relation for this eclipse
width and its error, calculated using equations~(\ref{eqn:incq})
and (\ref{eqn:eggleton}). These curves agree
to better than 1 per cent with those obtained using an exact computation
of the Roche-lobe geometry (see \pcite{dhillon92}). 
Even without a reliable estimate of the mass
ratio, we can constrain the 
inclination of SW Sex from Fig.~\ref{fig:inclrad} to be 
$i>75^\circ$. Note that \scite{penning84} derived a value 
of $i=79^\circ\pm1^\circ$, but this figure is based on a dubious radial 
velocity analysis (see section~\ref{sec:sysparams}).

The quasi-$B$-band continuum eclipses of \scite{rutten92b} have a mean
phase half-width at maximum light (ie. timing the first and
last contacts of eclipse and dividing by 2) of $\Delta\phi=0.12\pm0.02$
(Ren\'{e} Rutten, private communication). 
Following the method of \scite{sulkanen81}, it is possible to determine 
the radius of the accretion disc, $R_D$, in SW Sex through the geometric 
relation
\begin{equation}
R_D=a\sin{i}\sin{(2\pi\Delta\phi)}-R_C,
\label{eqn:chord1}
\end{equation}
where the half-chord on the secondary, $R_C$, is known in terms of
$i$ and $R_2/a$:
\begin{equation}
\frac{R_C}{a}=\sqrt{\left(\frac{R_2}{a}\right)^2-\cos^2{i}}.
\label{eqn:chord2}
\end{equation}

The lower panel of Fig.~\ref{fig:inclrad} shows a plot of the disc radius 
versus mass ratio
for three values of the eclipse width, $\Delta\phi=0.12\pm0.02$. 
The curves were produced by determining the inclination for each mass ratio
using equations~(\ref{eqn:incq}) and (\ref{eqn:eggleton}),
assuming $\Delta\phi_{1/2} = 0.098$, and
then calculating the disc radius for this inclination and mass ratio using
equations~(\ref{eqn:chord1}) and (\ref{eqn:chord2}). 
These curves agree to better than 5 per cent with those obtained using an 
exact computation of the Roche-lobe geometry (see \pcite{dhillon91}).
The dashed curve represents solutions where
$R_D<R_C$, i.e. solutions where we would expect to see a total 
(flat-bottomed) eclipse. This is clearly not the case for SW Sex, which 
exhibits round-bottomed eclipses (e.g. \pcite{rutten92b}). The accretion
disc in SW Sex must therefore have a radius which lies between 0.6$L_1$
(where $L_1$ is the distance between the white dwarf and the inner Lagrangian
point) and the radius of the primary Roche lobe, 
implying that the disc in SW Sex is
comparable in size to the discs of other nova-like variables and
dwarf novae in outburst, but is larger than the discs of dwarf novae
in quiescence (see \pcite{harrop96}). 

\subsection{Average spectrum}
\label{sec:avspec}

\begin{figure*}
\centerline{\psfig{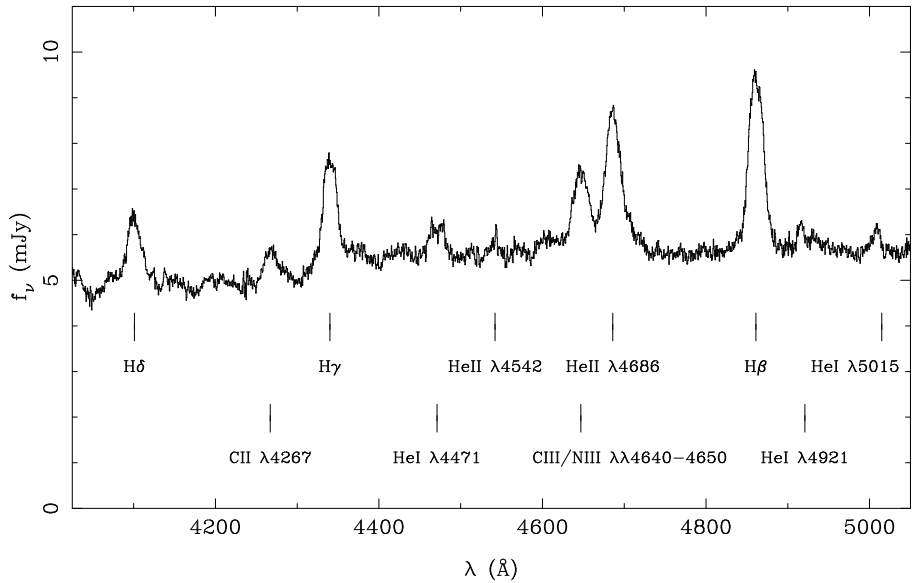}}
\caption{The average of all 158 spectra of SW Sex, uncorrected for
orbital motion.}
\label{fig:average}
\end{figure*}

\begin{table}
\caption{Fluxes, equivalent widths and velocity widths of
prominent lines in the average spectrum of SW Sex (Fig.~\ref{fig:average}). 
The errors
on the flux, EW, FWHM and FWZI measurements are of order 
0.1 erg\,cm$^{-2}$\,s$^{-1}$, 1\,\AA, 100 km\,s$^{-1}$ and 
1000 km\,s$^{-1}$, respectively.}
{\bf\boldmath
\begin{center}
\begin{tabular}{@{}lcrcc}
& & & & \\
\multicolumn{1}{l}{Line} &
\multicolumn{1}{c}{Flux} &
\multicolumn{1}{c}{EW} &
\multicolumn{1}{c}{FWHM} &
\multicolumn{1}{c}{FWZI} \\
& \multicolumn{1}{c}{$\times$10{\normalsize$^{\bf-13}$}} &
\multicolumn{1}{c}{\AA} &
\multicolumn{1}{c}{km\,s{\normalsize$^{\bf -1}$}} &
\multicolumn{1}{c}{km\,s{\normalsize$^{\bf -1}$}} \\
& \multicolumn{1}{c}{erg\,cm\,{\normalsize$^{\bf -2}$}\,s{\normalsize$^{\bf -1}$}} & & & \\
& & & & \\
H$\beta$                              & 1.22 & 17.3 & 1270 & 4500 \\
H$\gamma$                             & 1.09 & 13.3 & 1290 & 4800 \\
H$\delta$                             & 1.12 & 13.6 & 1450 & 4500 \\
HeI $\lambda$4471                     & 0.27 &  3.3 & 1500 & 3000 \\
HeI $\lambda$4921                     & 0.21 &  3.1 &      &      \\
HeI $\lambda$5015                     & 0.09 &  1.4 &      &      \\
HeII $\lambda$4542                    & 0.06 &  0.8 &      &      \\
HeII $\lambda$4686                    & 1.15 & 15.2 & 1400 & 5700 \\
CII $\lambda$4267                     & 0.24 &  2.9 & 1600 & 4500 \\
CIII/NIII                             & 0.77 &  9.9 & 1500 & 5000 \\
\end{tabular}
\end{center}
}
\label{tab:lines}
\end{table}

\begin{figure*}
\centerline{\psfig{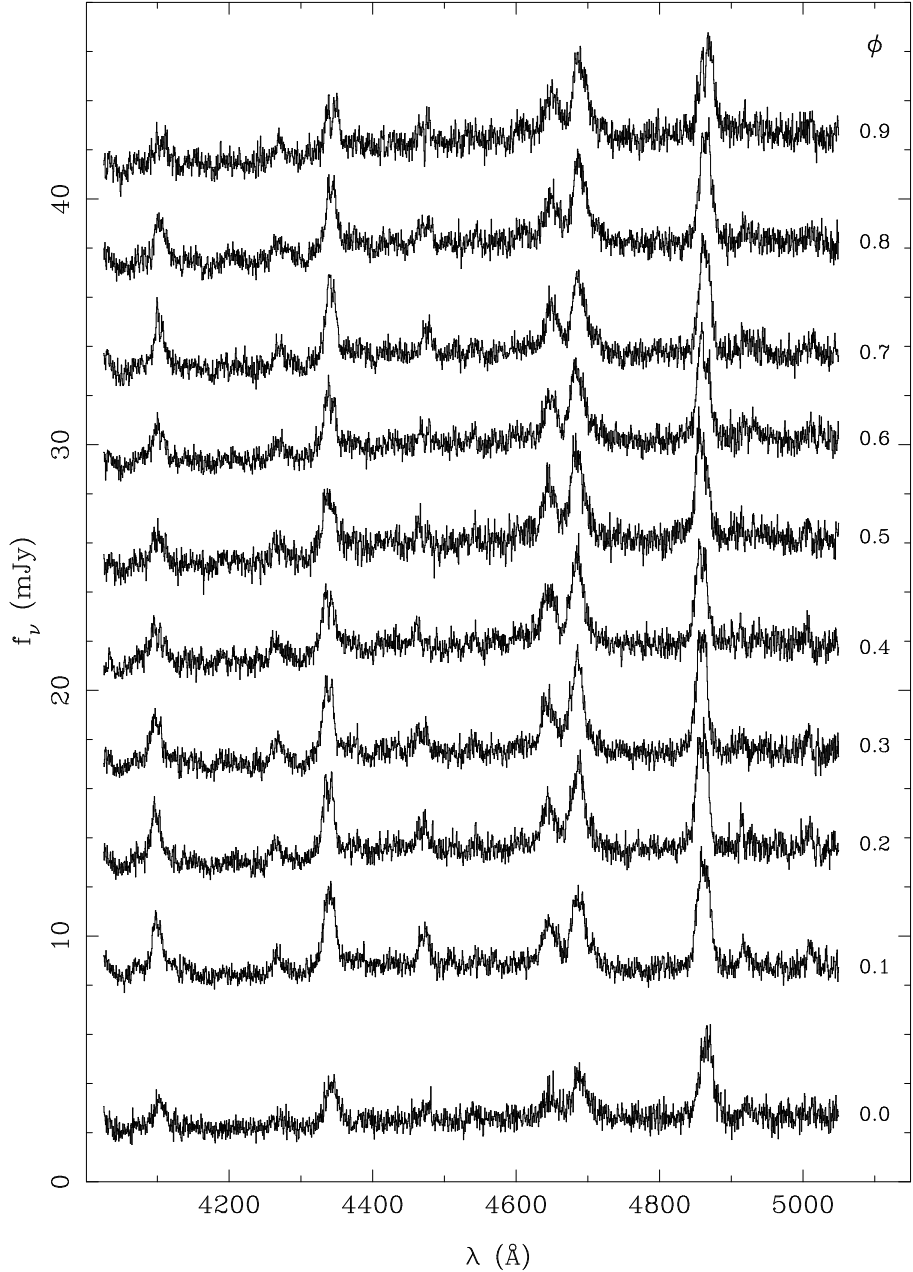}}
\caption{Orbital emission line variations in SW Sex. The data have been 
averaged into 10 binary phase bins, with a multiple of 4 added to each spectrum
in order to displace the data in the vertical direction.}
\label{fig:evol}
\end{figure*}

The average of all 158 spectra of SW Sex, uncorrected for orbital motion,
is displayed in Fig.~\ref{fig:average}. A direct comparison of this spectrum
with that of a number of other SW Sex stars is given in Fig.~5 of 
\scite{dhillon96}. In common with most other
high-inclination nova-like variables, including magnetic systems 
(e.g. FO Aqr; \pcite{marsh96}), SW Sex shows a number of features very 
different from those exhibited by high-inclination dwarf novae
(e.g. IP Peg; \pcite{marsh88a}). The low-excitation Balmer and 
He\,{\small I} lines show symmetric, single-peaked profiles instead of the 
double-peaked profiles one would expect from a high-inclination accretion 
disc. The high-excitation lines of
He\,{\small II} $\lambda$4686\,\AA, He\,{\small II} $\lambda$4542\,\AA, 
C\,{\small II} $\lambda$4267\,\AA\ and 
C\,{\small III/N\,{\small III}} $\lambda$4640--4650\,\AA\ are very strong,
much more prominent than in most quiescent dwarf novae, and also exhibit
symmetric, single-peaked profiles. In addition, there appear to be many
weak features in the continuum not usually seen in dwarf novae. 

In Table~\ref{tab:lines} we list fluxes, equivalent widths and velocity 
widths of the most prominent lines in the average spectrum of SW Sex. The 
Balmer decrement is flat, indicating that the emission is optically
thick, as is usually the case for cataclysmic variables.
The FWHM of the lines are comparable to those
observed in other SW Sex stars (e.g. V1315 Aql; \pcite{dhillon91}),
but are small compared to the FWHM of emission lines in high-inclination 
dwarf novae (e.g. IP Peg; \pcite{marsh88a}). The FWZI of the emission lines in 
SW Sex, however, are comparable with those observed in IP Peg. This 
indicates that the emission lines in SW Sex (and other members of its class)
have narrow cores and broad wings when compared with the profiles expected 
of a canonical high-inclination accretion disc. The average spectrum of 
SW Sex does differ slightly from the other members of its class in one 
respect -- the ratio of the strength of the high-excitation features 
to the strength of the Balmer lines is larger in SW Sex than is commonly 
observed in other SW Sex stars. 

\subsection{Emission line variations}
\label{sec:evol}

We rebinned the spectra on to a uniform wavelength scale and cast the
data into 10 binary phase bins by averaging all the spectra falling into
each bin. A multiple of 4 was then added to each spectrum in order to
displace the data along the ordinate. The result is plotted in 
Fig.~\ref{fig:evol}. 

The Balmer line-profile variations are extremely complex and very 
similar to those observed in other SW Sex stars 
(e.g. DW UMa, \pcite{shafter88}; V1315 Aql, \pcite{dhillon91}). 
The lines are single peaked, but show transient
absorption features at phases 0.2--0.4 and 0.6--0.9, giving the profiles a
double-peaked appearance. The 
absorption features in SW Sex are not as deep as those in V1315 Aql and they 
appear to be visible for a greater fraction of the orbit. 
The He\,{\small I} lines, and to a lesser extent the C\,{\small II}
$\lambda$4267\,\AA\ line, 
show absorption features which behave in a similar way to the 
Balmer lines. The lines of He\,{\small II} 
$\lambda$4686\,\AA\ and
C\,{\small III}/N\,{\small III} $\lambda$4640--4650\,\AA, 
on the other hand, appear to be single
peaked for the entire orbit. The spectrum at the foot of
Fig.~\ref{fig:evol} shows that the lines are relatively unchanged by 
primary eclipse, in approximate agreement with the behaviour of other SW 
Sex stars. Note, however, that He\,{\small II} $\lambda$4686\,\AA\ 
and C\,{\small III}/N\,{\small III} $\lambda$4640--4650\,\AA\ 
appear to be more deeply eclipsed in V1315 Aql than in SW Sex. 

\begin{figure*}
\centerline{\psfig{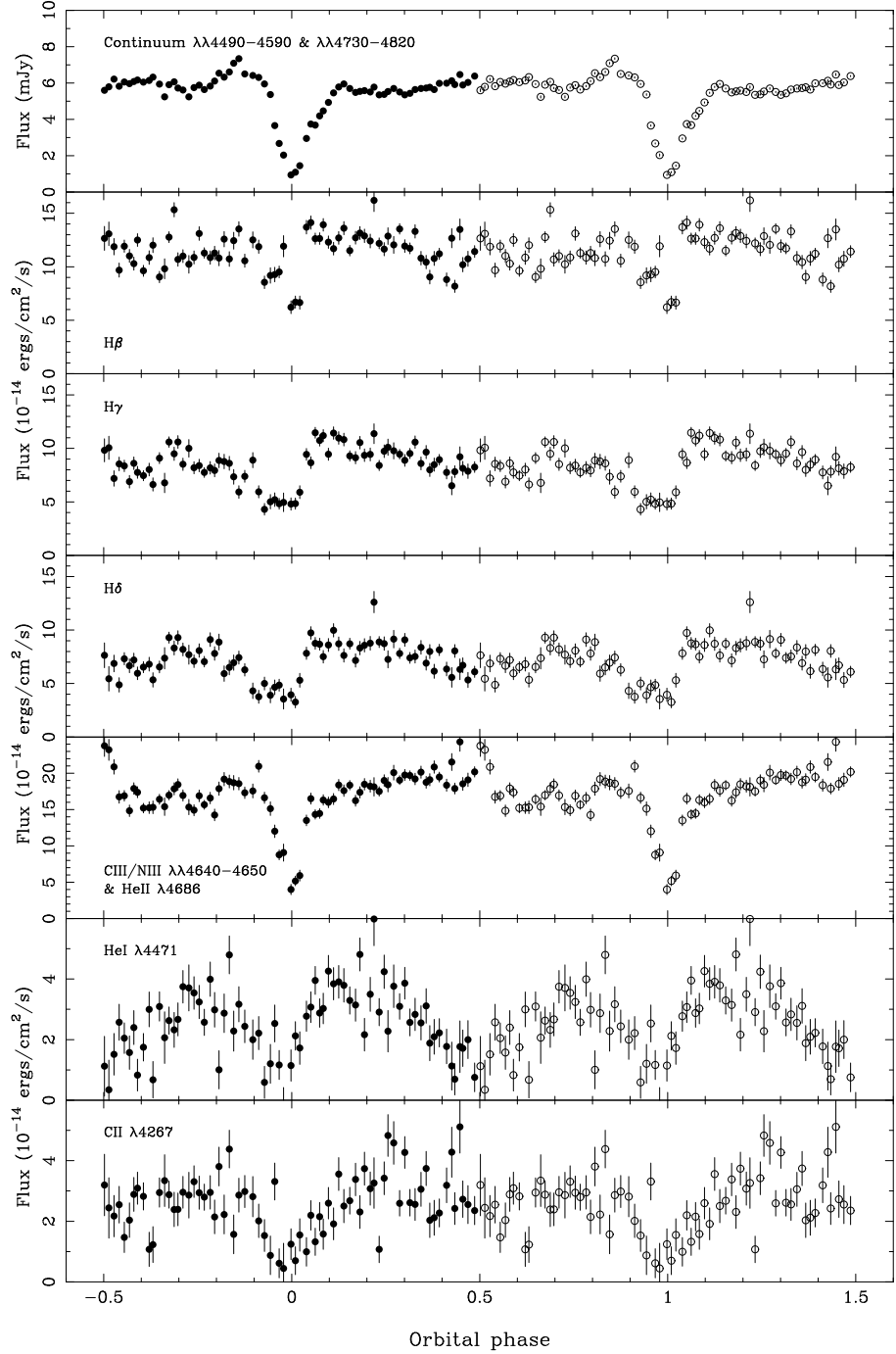}}
\caption{Continuum and line light curves of SW Sex. The open circles
represent points where the real data (closed circles) have been folded
over.}
\label{fig:lc}
\end{figure*} 

\subsection{Light curves}
\label{sec:lc}

We binned the data into 75 phase bins and fitted the continuum of
each spectrum with a third-order polynomial. The flux in each line was then
derived by summing the residual flux in the line after continuum 
subtraction. Fig.~\ref{fig:lc} shows the continuum light curve
(measured between 4490--4590\,\AA\ and 4730--4820\,\AA) and those of
H$\beta$, H$\gamma$, H$\delta$, He\,{\small I} $\lambda$4471\,\AA, 
C\,{\small II} $\lambda$4267\,\AA\ and the sum 
of He\,{\small II} $\lambda$4686\,\AA\ and 
C\,{\small III/N\,{\small III}} $\lambda$4640--4650\,\AA. Note that the
data have been folded so as to display more than one binary cycle. 

A direct comparison of these light curves with those of a number of
other SW Sex stars is given in Fig.~5 of \scite{dhillon96}.
The continuum light curve shows a deep, rounded eclipse, with a small
hump before primary eclipse and a shoulder during egress. \scite{ashoka94} 
observed a similar hump prior to eclipse on three consecutive orbits of 
SW Sex, thereby ruling out the possibility that the humps are due to 
flickering. The origin of these features is almost certainly due to the 
changing aspect of the bright-spot and is supported by the appearance of the
strong bright-spot, contributing 13 per cent of the total light, in the 
eclipse maps of SW Sex presented by \scite{rutten92b}. 

The Balmer-line light curves are remarkable. They are not 
as deeply eclipsed as the continuum and show wide, asymmetric 
eclipse profiles with a broad ingress and very steep egress. 
The ingress is particularly wide in H$\gamma$ and H$\delta$,
implying that some of the line flux is eclipsed as early as 
phase $\sim0.8$, which is too early to be explained in terms
of obscuration by the secondary star. 
The eclipses in H$\gamma$ and H$\delta$ appear
to reach a minimum before phase 0 and show flat-bottomed profiles. 
The egress in these lines is extremely rapid, commencing 
around phase 0.01 and completing by phase 0.05. 
There is also evidence for a reduction in line flux around phase 
0.5 in all of the Balmer lines, as observed so prominently in other 
SW Sex stars, and this feature becomes stronger as one moves up 
the Balmer series. The He\,{\small I} $\lambda$4471\,\AA\ line
shows a similar eclipse profile to H$\delta$, but the eclipse is
deeper and the phase 0.5 absorption much stronger. The 
C\,{\small II} $\lambda$4267\,\AA\ line shows an eclipse shape which
is reversed to that of the other lines, with a narrow ingress and
a broad egress which finishes at phase $\sim0.2$ (too late to be 
explained in terms of obscuration by the secondary star).  
The He\,{\small II} $\lambda$4686\,\AA\ and 
C\,{\small III/N\,{\small III}} $\lambda$4640--4650\,\AA\ light curve is
the only one which comes close to resembling the continuum light curve -- 
the eclipse is deep and approximately of the same width as the continuum
eclipse, although less rounded, and appears to be preceded by a
small orbital hump. 

\begin{figure*}
\centerline{\psfig{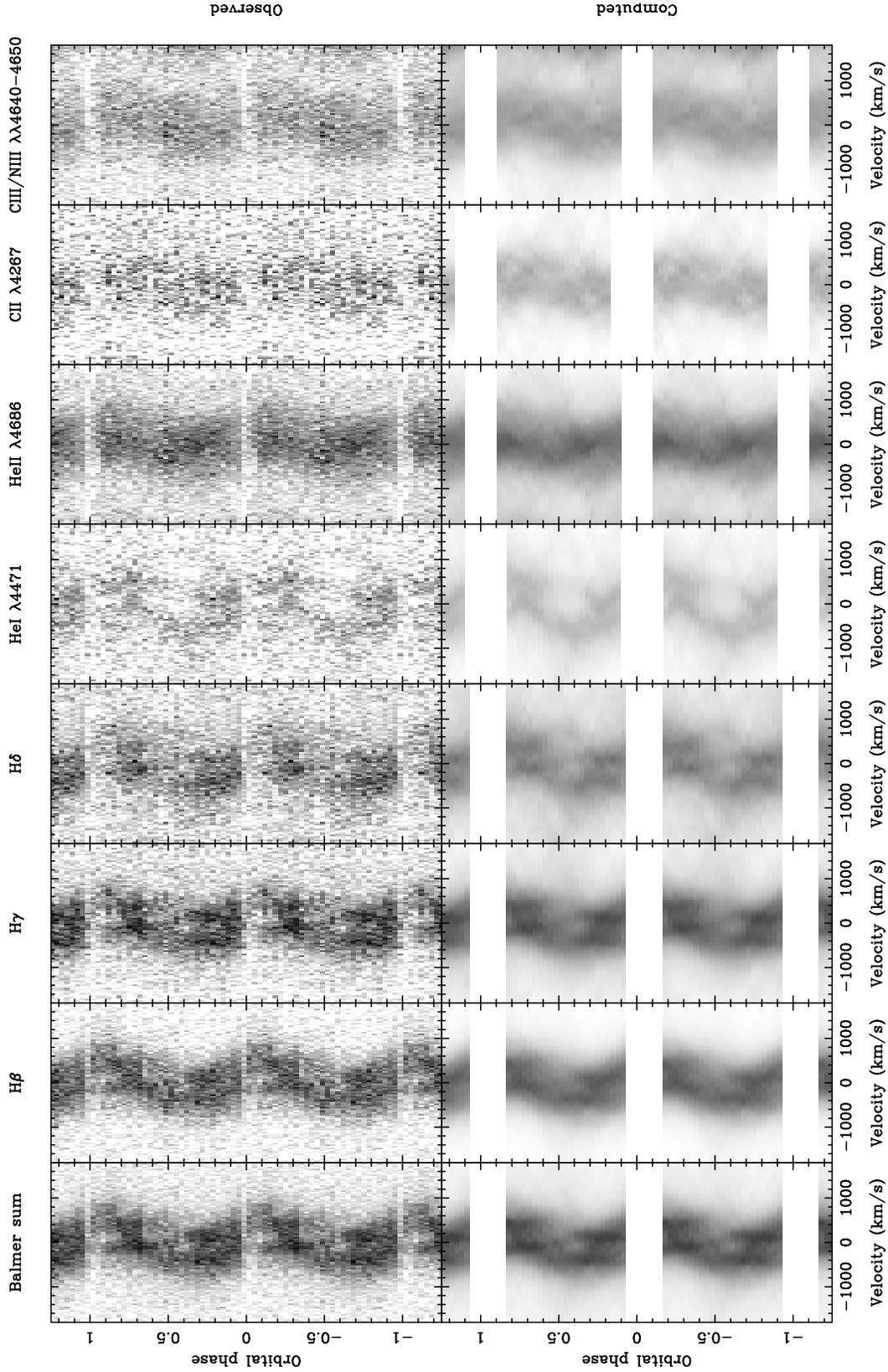}}
\caption{Upper panel -- trailed spectra of the sum of the Balmer
lines, H$\beta$, H$\gamma$, H$\delta$, He\,{\small I} $\lambda$4471\,\AA,
He\,{\small II} $\lambda$4686\,\AA, C\,{\small II} $\lambda$4267\,\AA\ and
C\,{\small III/N\,{\small III}} $\lambda$4640--4650\,\AA\ in SW Sex. 
Lower panel -- computed data from the Doppler maps 
of the above emission lines presented in \protect{Fig.~\ref{fig:maps}.}
The gaps correspond to eclipse spectra which were omitted 
from the fit. Note that the data have been folded in order to display 
more than one binary cycle.}
\label{fig:trail}
\end{figure*} 

\subsection{Trailed spectra}
\label{sec:trail}

We binned the data into 30 binary phase bins and subtracted the
continuum from each spectrum using a third-order polynomial fit. We then
rebinned the data on to a constant velocity interval scale of 35 
km\,s$^{-1}$\,pixel$^{-1}$, centred on the H$\beta$, H$\gamma$, H$\delta$,
He\,{\small I} $\lambda$4471\,\AA, He\,{\small II} $\lambda$4686\,\AA, 
C\,{\small II} $\lambda$4267\,\AA\ and
C\,{\small III/N\,{\small III}} $\lambda$4640--4650\,\AA\ lines. The
trailed spectra of these lines are shown in the upper panel of
Fig.~\ref{fig:trail}, along with the trailed spectra of the sum of the
Balmer lines (computed by taking the weighted average of the three Balmer 
lines H$\beta$:H$\gamma$:H$\delta$ as 0.41:0.35:0.24). Note that the data
have been folded so as to display more than one binary cycle. 

The trailed spectra of SW Sex appear to show a number of clearly-defined
components. The most dominant is a strong orbital modulation with a 
semi-amplitude of $\sim500$~km\,s$^{-1}$ which crosses zero velocity from
red to blue at phase $\sim$0.15. This feature appears to be present in all
of the emission lines (in the high-excitation features it appears to be the
only component) 
and shares the velocity and phasing of what one might
expect of the bright spot. 
There is a second, lower-velocity component which
can only be made out in the Balmer lines, and is particularly well-defined
in the plot of the sum of the Balmer lines. This modulation has a 
semi-amplitude of $\sim200$~km\,s$^{-1}$ and appears to be 
in approximate anti-phase to the dominant modulation, crossing zero velocity
from blue to red at phase $\sim$0. The most likely origin of this emission
is, therefore, the secondary star. There is also evidence for a third component
in the Balmer lines, which is most readily apparent at a velocity of 
$\sim400$~km\,s$^{-1}$ at phase 0.75 and which appears to be associated
with what might be a faint double-peaked accretion disc component. 
This component is also visible in He\,{\small I} $\lambda$4471\,\AA.
Note that the three emission components we have identified in the trailed
spectra all appear to be eclipsed by similar amounts around phase
0 and attenuated by similar amounts around phase 0.5. A further
discussion of the various emission regions in SW Sex is deferred until
section~\ref{sec:doptom}, where we report on the results of our
Doppler tomography experiments.

\subsection{Radial velocities}
\label{sec:rv}

\begin{figure}
\centerline{\psfig{figure=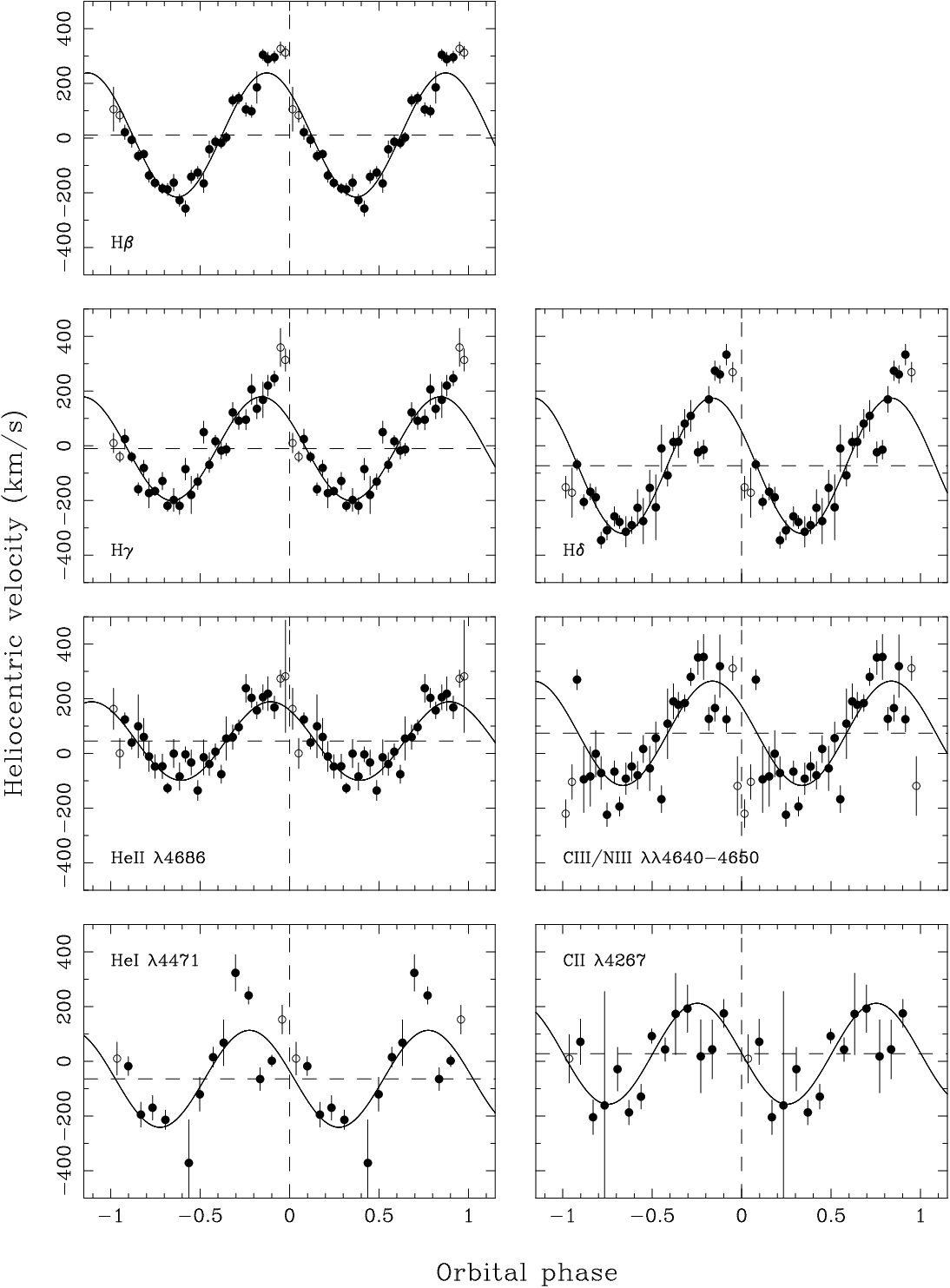,width=8.4cm}}
\caption{Radial velocity curves of H$\beta$, H$\gamma$, H$\delta$,  
He\,{\small II} $\lambda$4686\,\AA,
C\,{\small III/N\,{\small III}} $\lambda$4640--4650\,\AA, 
He\,{\small I} $\lambda$4471\,\AA\ and C\,{\small II} $\lambda$4267\,\AA\ 
in SW Sex, measured using a double-Gaussian separation of 1600 km\,s$^{-1}$.
Points marked by open circles were not 
included in the fits due to measurement uncertainties during primary 
eclipse. The horizontal dashed lines represent the systemic velocities.
Note that the data have been folded in order to display more than 
one binary cycle.}
\label{fig:rv}
\end{figure} 

A measure of the radial velocity semi-amplitudes of either or both the 
white dwarf and secondary star is essential in order to derive the 
component masses of SW Sex. 
We measured radial velocities of the emission lines in SW Sex by 
applying the double-Gaussian method of \scite{schneider80} to the
data presented in the upper panel of Fig.~\ref{fig:trail}. 
This technique is most 
sensitive to the motion of the line wings and should thus reflect the 
radial velocity semi-amplitude of the white dwarf, $K_W$, with the 
highest reliability. The Gaussians 
were of width 200 km\,s$^{-1}$ ($\sigma$) and we varied their separation 
from 600 to 2200 km\,s$^{-1}$. We then fitted 
\begin{equation}
V=\gamma-K\sin(\phi-\phi_0)
\end{equation}
to each set of measurements, where $V$ is the radial velocity, 
$K$ is the semi-amplitude, $\phi$ is the orbital phase and $\phi_0$ is the
phase where the radial velocity curve crosses from red to blue. 
We omitted four points around phase 0 during the fitting procedure owing to 
measurement uncertainties during primary eclipse. 

\begin{figure*}
\centerline{\psfig{figure=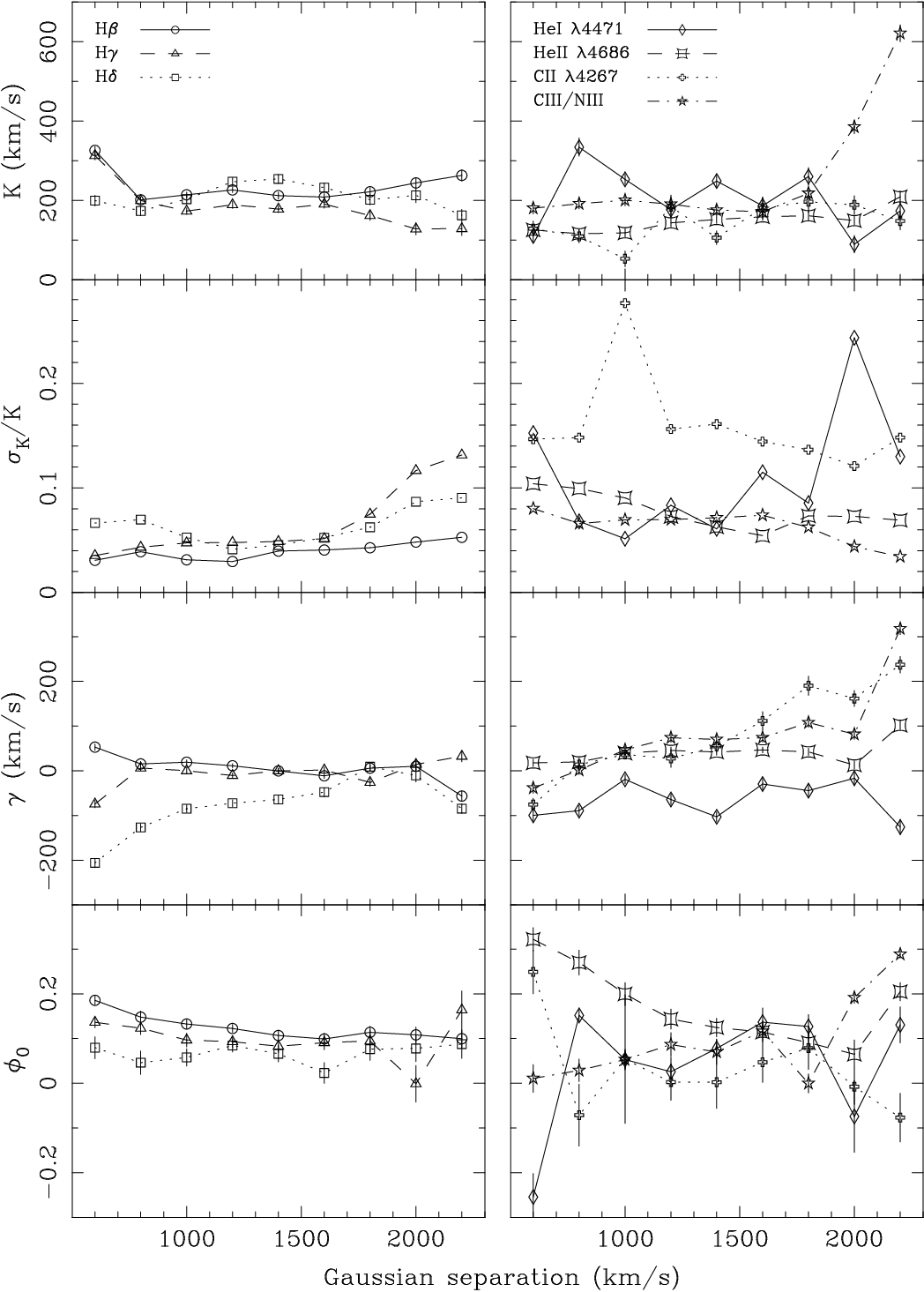,width=12cm}}
\caption{Diagnostic diagrams for SW Sex. 
The left-hand panel shows the results of the
radial velocity fits to 
H$\beta$ (circles connected by solid lines), H$\gamma$ (triangles
connected by dashed lines) and H$\delta$ (squares connected by dotted lines).
The right-hand panel shows the results of the radial velocity fits 
to He\,{\small I} $\lambda$4471\,\AA\ (diamonds connected by solid lines),
He\,{\small II} $\lambda$4686\,\AA\ (squares connected by dashed lines),
C\,{\small II} $\lambda$4267\,\AA\ (crosses connected by dotted lines),
and C\,{\small III/N\,{\small III}} $\lambda$4640--4650\,\AA\  
(stars connected by dashed-dotted lines).}
\label{fig:diagn}
\end{figure*} 

Examples of the radial velocity curves obtained using Gaussian separations 
of 1600~km\,s$^{-1}$ are shown in Fig.~\ref{fig:rv}. 
The most striking feature exhibited by these radial velocity curves 
are the phase shifts, where 
(with the exception of the relatively noisy 
He\,{\small I} $\lambda$4471\,\AA\ and C\,{\small II} $\lambda$4267\,\AA\ 
curves)
the spectroscopic conjunction of each line, 
i.e. the superior conjunction of the emission line source, occurs after 
photometric conjunction, i.e. photometric mid-eclipse. 
This phase shift 
implies an emission line source trailing the accretion disc, such as the 
bright spot, and has been observed in other SW Sex stars (e.g. 
DW UMa, \pcite{shafter88}; V1315 Aql, \pcite{dhillon91}). There is also
some evidence of a rotational disturbance in the emission 
lines of SW Sex during eclipse, as it can be seen that radial 
velocities measured just prior to eclipse are skewed to the red and those 
measured just after eclipse are (arguably) skewed to the blue. This confirms 
the detection of a similar feature in the trailed spectra of 
Fig.~\ref{fig:trail} and indicates that at least part of the line emission 
may originate in an accretion disc. 

The final results of the radial velocity analysis are displayed in the
form of a diagnostic diagram \cite{shafter86} in Fig.~\ref{fig:diagn}. 
By plotting $K$, its associated fractional error $\sigma_K/K$, $\gamma$
and $\phi_0$ as functions of the Gaussian separation, it is possible to 
select the value of $K$ which most closely matches $K_W$. 
If there is disc emission contaminated by bright-spot
emission, for example, one would expect the solution for $K$ to 
asymptotically approach $K_W$ when the Gaussian separation becomes sufficiently
large. Furthermore, since the phasing of the bright-spot is not the same
as the white dwarf, one would expect $\phi_0$ to approach 0 as the
Gaussian separation increases. In order to determine the maximum useful 
value of the Gaussian separation, one can inspect the fractional error 
($\sigma_K/K$) curve for an increase, which indicates that the 
measurements are beginning to be dominated by noise. Taking the results
for the Balmer lines in SW Sex first (left-hand panel of Fig.~\ref{fig:diagn}),
the point where the fractional error starts increasing is at a Gaussian
separation of 1600~km\,s$^{-1}$. This means that $K$ is always greater than
$\sim200$~km\,s$^{-1}$, regardless of the Gaussian separation. As pointed out
by \scite{shafter88}, who noted a similar value in DW UMa, this value is
distressingly high, implying a very low white dwarf mass. It is clear from
the $\phi_0$ curves, however, that these $K$-values cannot possibly
represent the white dwarf, as even at large Gaussian separations 
$\phi_0\sim0.1$, implying that nearly all of the line emission must originate
from a source, such as the bright-spot, which trails the accretion disc. 
Clearly, one cannot adopt a reliable value of $K_W$ from an analysis of the 
Balmer emission. The same can also be said for the higher-excitation species
displayed in the right-hand panel of
Fig.~\ref{fig:diagn}, which behave in a largely similar fashion to the 
Balmer lines, except that they seem to show lower $K$-velocities and larger 
$\gamma$-velocities. Note that we also attempted to use the 
light centre technique of \scite{marsh88a} to determine $K_W$, but do not
report on the results here as the method was clearly unsuitable to this
dataset (as was also found to be the case for V1315 Aql; \pcite{dhillon91}). 

\subsection{Doppler tomography}
\label{sec:doptom}

The velocity profile of a line observed at a given binary phase is simply
the projection of the system's velocity field on to the line of sight. The
orbital motion of the system provides projections along different lines of
sight, and thus the velocity field can be recovered from the trailed spectra. 
This technique is known as Doppler tomography. Full technical details of
the method are given by \scite{marsh88b} and \cite{horne91} and an example of 
its application to real data is given by \scite{kaitchuck94}. 

\begin{figure*}
\centerline{\psfig{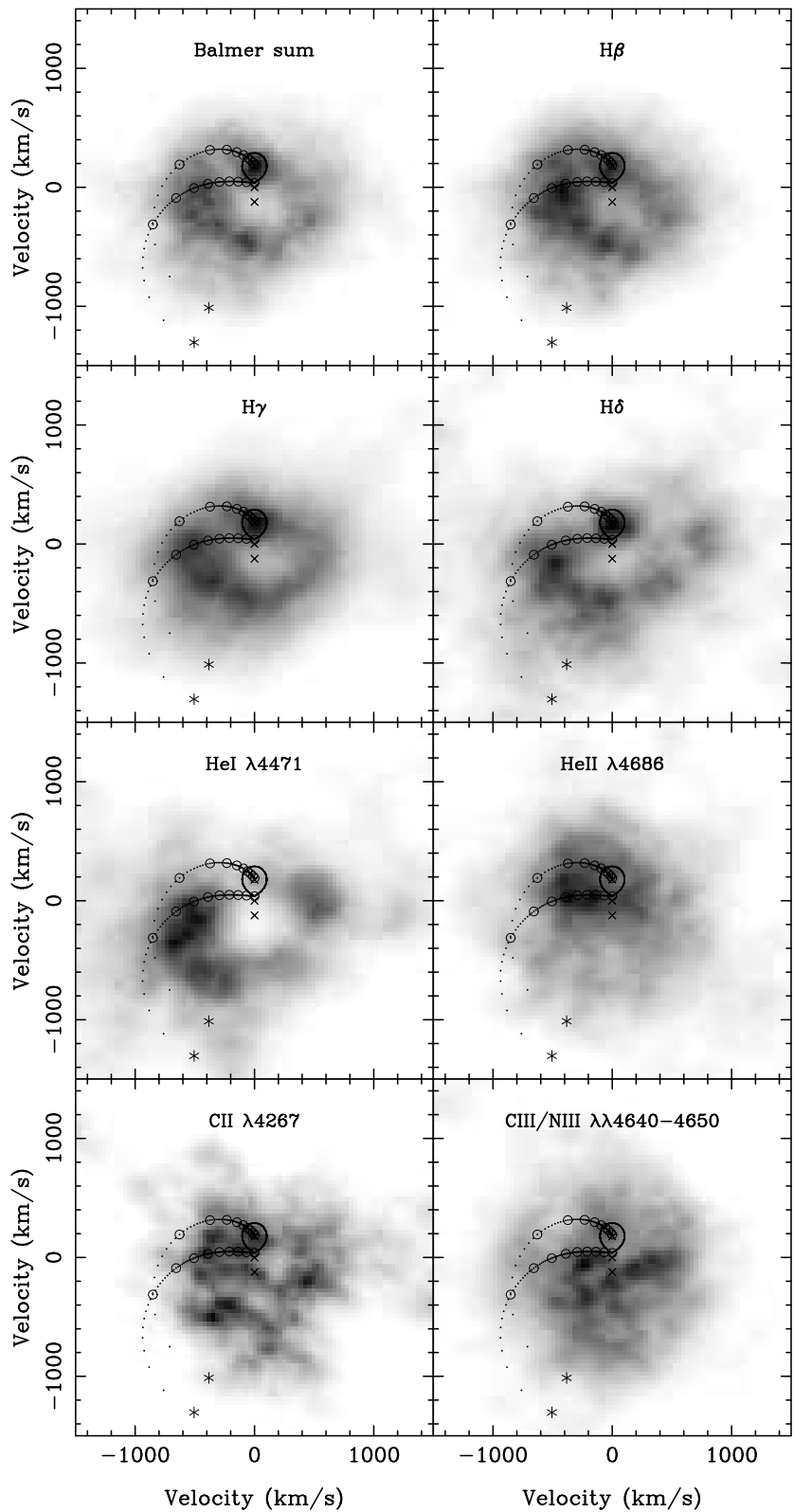}}
\caption{Doppler maps of the sum of the Balmer lines, H$\beta$, H$\gamma$, 
H$\delta$, He\,{\small I} $\lambda$4471\,\AA, He\,{\small II} 
$\lambda$4686\,\AA, C\,{\small II} $\lambda$4267\,\AA\ and
C\,{\small III/N\,{\small III}} $\lambda$4640--4650\,\AA\ 
in SW Sex. The three crosses in each map are, from top to bottom, the
centres of mass of the secondary, the system and the white dwarf. The
predicted outline of the secondary star is marked along with the  
gas stream and the Keplerian velocity of the disc along the gas stream 
(the lower and upper curves, respectively).
The series of small circles along the gas streams mark the distance from the
white dwarf at intervals of $0.1L_1$, ranging from $1.0L_1$ at the 
secondary star to the point of closest approach (marked by an asterisk).}
\label{fig:maps}
\end{figure*} 

Fig.~\ref{fig:maps} shows the Doppler maps of 
the sum of the Balmer lines, H$\beta$, H$\gamma$, 
H$\delta$, He\,{\small I} $\lambda$4471\,\AA, He\,{\small II} 
$\lambda$4686\,\AA, C\,{\small II} $\lambda$4267\,\AA\ and
C\,{\small III/N\,{\small III}} $\lambda$4640--4650\,\AA.
The maps have been computed from the trailed spectra presented in the 
upper panel of Fig.~\ref{fig:trail}, with the exception of 
He\,{\small II} $\lambda$4686\,\AA\ and C\,{\small III/N\,{\small III}} 
$\lambda$4640--4650\,\AA, which are blended and were thus computed together 
(see \pcite{marsh88b}). The three crosses marked 
on each Doppler map represent the centres of mass of the secondary (upper 
cross), the system (central cross) and the primary (lower cross). The
predicted trajectory of the gas stream and the Keplerian velocity
of the disc along the gas stream have also been plotted 
(the lower and upper of the two curves, respectively), along 
with the secondary's Roche lobe. These have been computed using 
representative, but realistic, system parameters of 
$q=0.7$ and $K_R+K_W=300$~km\,s$^{-1}$ (see section~\ref{sec:sysparams}). 
The series of small circles along the gas streams mark the distance from the
white dwarf at intervals of $0.1L_1$, ranging from $1.0L_1$ at the 
secondary star to the point of closest approach (marked by an asterisk). 
The lower panel in Fig.~\ref{fig:trail} shows the
data computed from the Doppler maps of Fig.~\ref{fig:maps}; 
a comparison of the upper and lower
panels of Fig.~\ref{fig:trail} gives some indication of the quality of the
fit. The gaps in the computed data correspond to eclipse spectra
(defined for each line by the shape of the eclipse light curve in
Fig.~\ref{fig:lc}) which have been omitted from the fit as Doppler tomography 
cannot properly account for obscuration of line-emitting material. 

The Balmer-line Doppler maps appear to be dominated by emission from three 
components -- the secondary star, the accretion disc and an extended
bright spot.
The accretion disc manifests itself as a ring-like structure in the maps,
centred on the white dwarf and with a radius of $\sim500$~km\,s$^{-1}$. If
this velocity represents the outer disc velocity, then the radius of the
accretion disc would be $R_D\sim0.6 L_1$ (assuming a representative,
but realistic, mass ratio of $q=0.7$; see
section~\ref{sec:sysparams}), in good agreement with 
the photometric determination of the disc radius presented in 
section~\ref{sec:inclrad}. Superposed on the ring-like disc emission in
the Balmer-line maps is a bright spot which has a velocity close to
(but not exactly equal to) the expected velocity of the gas stream at a
distance from the white dwarf of $\sim0.5L_1$, in reasonable agreement with
the disc radius determinations just described. Of course, the precise
location of this emission region with respect to the gas stream depends
on the system parameters adopted. The bright spot
appears much more strongly in the H$\beta$ map than in the H$\gamma$ and
H$\delta$ maps and gives the visual impression of line emission 
originating from the bright spot but extending downstream along the disc edge.
The Balmer line maps also display prominent emission from the secondary star,
which becomes more dominant relative to the bright
spot and disc emission as one moves from H$\beta$ to H$\delta$. 
The map of He\,{\small I} $\lambda$4471\,\AA\ appears similar to the
Balmer-line maps but shows no evidence of secondary star emission. 
The map of He\,{\small II} $\lambda$4686\,\AA, appears to show only one 
emission-line source, located with a velocity equal to the
gas stream at a distance from the white dwarf of $\sim0.5L_1$ and hence 
also most likely associated with the bright spot. There is also
some evidence that the emission is extended along the gas stream to even
greater distances from the white dwarf. Whereas the Balmer-line emission from 
the bright-spot is not quite coincident with the velocity of the gas stream
(assuming our representative system parameters), the He\,{\small II} 
$\lambda$4686\,\AA\ emission lies almost exactly on the expected velocity of 
the stream and shows no evidence for an extension along the disc rim.

\subsection{System parameters}
\label{sec:sysparams}

\begin{figure*}
\centerline{\psfig{figure=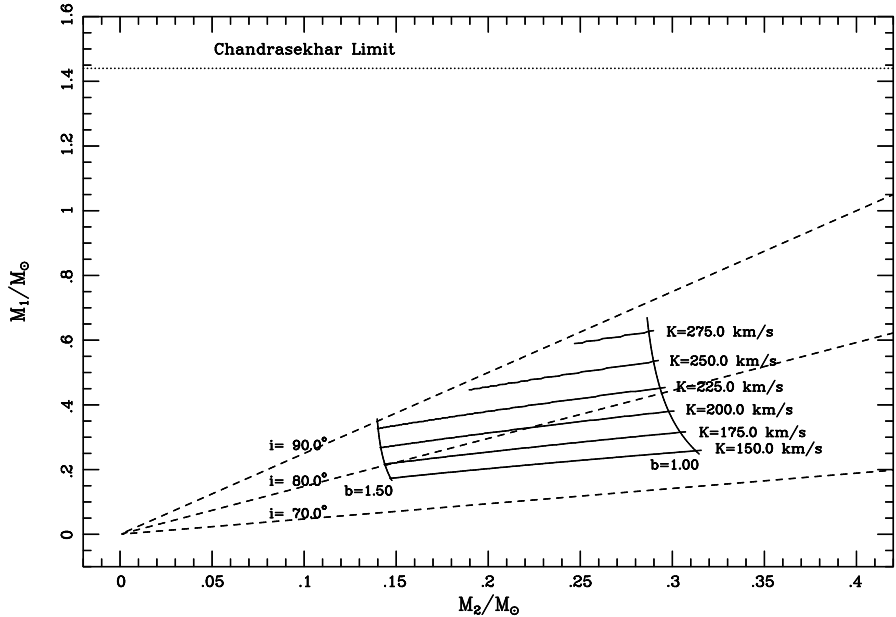,width=14cm,rheight=8.5cm}}
\caption{Constraints on the component masses of SW Sex. The white 
dwarf mass is plotted as a function of the secondary star mass for 
representative values of $K_R$ and the mass-radius relation parameter
$b$. Loci of constant inclination (and constant mass ratio) are shown 
as dashed lines. The horizontal dotted line indicates the 
Chandrasekhar limit on the mass of the white dwarf. The lines marked 
with $b=1$ and $b=1.5$ represent secondaries having a mean density 
near the main sequence and less than the main sequence, respectively.}
\label{fig:sysparams}
\end{figure*} 

The radial velocity analysis described in section~\ref{sec:rv} and
the Doppler maps presented in Fig.~\ref{fig:maps} highlight the dangers 
of using emission lines in CVs to determine the motion of the white 
dwarf due to the presence of contaminating emission sources. 
The system parameters of SW Sex derived in this manner by \scite{penning84},
$M_1=0.58\pm0.20 M_{\odot}$ and $M_2=0.33\pm0.06 M_{\odot}$,
must therefore be viewed as highly dubious. 
We are fortunate in SW Sex, however, to have detected emission 
from the secondary star in the Doppler maps. This emission is most
likely produced by irradiation of the inner face of the secondary by
the accretion disc or bright-spot, as has been observed in 
DW UMa (\pcite{dhillon94}; \pcite{rutten94}). As the radial velocity
of the inner face of the secondary 
about the centre of mass of the system will be lower than
the radial velocity of the centre of mass of the secondary, a measurement
of the velocity of the emission feature from the Doppler map can be used
to set a lower limit to the true radial velocity semi-amplitude of the
secondary, $K_R$. From the Doppler map of the sum of the Balmer lines 
shown in Fig.~\ref{fig:maps}, we find $K_R>180$~km\,s$^{-1}$. 

We can now place limits upon the system parameters of SW Sex. Following
the method outlined by \scite{shafter84}, it is
necessary to solve two $q$ versus $i$ relations using our derived
values of $K_R$, $P$ and $\Delta\phi_{1/2}$. One $q$ versus $i$
relation is obtained by combining equations~(\ref{eqn:incq}) and
(\ref{eqn:eggleton}). The second $q$ versus $i$ relation is derived by
combining the mass function of the secondary star,
\begin{equation}
M_{2} = q^{3}(1+1/q)^{2}PK_{R}^{3}/(2\pi G\sin^3{i}),
\label{eqn:massfn}
\end{equation}
with Kepler's third law and a power-law mass-radius relation for
the secondary of the form
\begin{equation}
R_{2}/R_{\odot} = b(M_{2}/M_{\odot})^{x},
\label{eqn:kepler}
\end{equation}
where the parameters $b$ and $x$ have been determined empirically
by \scite{patterson84}, who finds that $b=1$ and $x=0.88$ for the
lower main-sequence. 

The results of the preceeding analysis applied to SW Sex are 
shown in Fig.~\ref{fig:sysparams}. The white dwarf mass is plotted
as a function of the secondary star mass for various values of
$K_R$ and the mass-radius relation parameter $b$ (adopting 
$x=0.88$, $\Delta\phi_{1/2}=0.098$ and $P=0.1349384229$~d). Loci
of constant inclination (and constant mass ratio) are shown as
dashed lines. The horizontal dashed line indicates the 
Chandrasekhar limit on the mass of the white dwarf. The lines
marked with $b=1$ and $b=1.5$ represent secondaries having mean
densities near the main sequence and less than the main sequence,
respectively. It is clear from Fig.~\ref{fig:sysparams} that,
without a reliable estimate of $K_R$ and the mass-radius relation
of the secondary, it is impossible to obtain tight constraints on 
the masses. By adopting the limit $K_R>180$~km\,s$^{-1}$, however, 
we can say that the secondary mass must be below $\sim0.3M_{\odot}$ and 
the primary mass must lie in the range $\sim 0.3-0.7 M_{\odot}$. 
If the secondary is less dense than a main-sequence 
star, these limits become more severe. 

\section{Discussion} 
\label{sec:discussion} 

The literature on the SW Sex phenomenon has been rapidly expanding in recent 
years. The original theory of an accretion disc wind (\pcite{honeycutt86}) and
its variants (\pcite{dhillon95b}; \pcite{murray96}) have been followed by 
theories of Stark broadening \cite{lin88}, magnetically driven inflows
\cite{williams89}, outflows (\pcite{tout93}; \pcite{wynn95}) and bright-spot 
overflows (\pcite{hellier94}; \pcite{hellier96}). Judging by the results of 
our Doppler tomography experiments (section~\ref{sec:doptom}), it seems that, 
in the case of SW Sex at least, a much more simple model will suffice. We have 
identified three emission components, the secondary star, accretion disc and 
an extended bright spot. In this model, it is the dominance of single-peaked 
line emission from the bright spot over the weak double-peaked disc 
emission which gives SW Sex its single-peaked profiles. It is also this 
dominance which forces the radial velocity curves to follow the motion of the 
bright spot and thus exhibit large phase shifts. The transient absorption 
features in the Balmer-line profiles are mostly artifacts of the complex 
intertwining of the emission components from the secondary star,
bright spot and accretion disc and involve little true absorption. 
While we are confident of the accretion disc and secondary star
components of this model, we are less certain of the dominant
bright-spot component as it appears to be inconsistent with the
emission-line light curves. We discuss why this is the case in 
section~\ref{sec:conclusions}, but turn first to a more detailed
description of the three emission components of our model. 

\subsection{Balmer emission from the secondary star}

Emission from the inner face of the secondary in nova-likes is not uncommon. 
To date, it has been seen in MV Lyr \cite{schneider81}, 
TT Ari \cite{shafter85}, IX Vel \cite{beuermann90}, 
RW Sex \cite{beuermann92b}, DW UMa (\pcite{dhillon94}; \pcite{rutten94}), 
UX UMa \cite{kaitchuck94}, RW Tri \cite{still95a}, 
PG\,0859+415 \cite{still96} and V347 Pup \cite{still97}. 
Its detection
in SW Sex is unusual in that it appears to have been detected while SW Sex
was in a normal state (judging by the flux levels in our data when compared
with those of \pcite{still95}), whereas most previous detections of the 
secondary in nova-likes typically occur during periods of low (e.g. DW UMa) or 
high (e.g. RW Tri) mass transfer. Furthermore, observations of SW Sex
obtained two years after the observations presented in this paper
also show evidence for secondary star emission \cite{marsh97},
implying that this is probably not a transient feature. The most likely cause 
of the emission is irradiation of the inner hemisphere of the secondary by 
ultraviolet light from the bright spot, accretion disc or 
boundary layer, although chromospheric emission from the rapidly rotating, 
late-type secondary cannot be ruled out. With higher signal-to-noise data
it should be possible to discriminate between these
sources of emission by searching for asymmetries in the 
distribution of the Balmer emission on the surface of the secondary as
one might expect, for example, if the emission were due to irradiation by the
bright-spot. As there is no evidence for any
He\,{\small II} emission from 
the secondary we may conclude that soft X-rays from the boundary layer are 
either not produced or are absorbed by the disc. It should be noted that
irradiation of the secondary star in CVs is now seen as a possible 
contributor to the long-term evolution of CVs (see \scite{kolb96} and 
references therein).

\subsection{Balmer and He\,{\small I} emission from the accretion disc}

Unambiguous evidence for line emission from the discs of nova-like 
variables is not as common as one might think. Few nova-likes show
double-peaked Balmer or He\,{\small I} emission lines and, to the
best of our knowledge, no nova-likes show double-peaked He\,{\small II}
$\lambda$4686\,\AA\ emission. Similarly, few high-inclination
nova-likes show clear evidence for a rotational disturbance during
primary eclipse in any of the emission lines. One possible reason for
this is that the high mass transfer rates of nova-likes inferred from
eclipse maps imply that the discs are largely optically thick in the continuum
(\pcite{rutten92b}; \pcite{williams80}).
Balmer line-emission from the disc will then come from a relatively 
small region at the edge of the disc which is optically thin in the continuum, 
resulting in lines which are narrow, double-peaked and weak. 
This is precisely what is observed in the disc component of SW Sex
and probably accounts for the rarity of the disc component in other nova-likes
(where emission from the $(-V_x,-V_y)$ quadrant of Doppler maps invariably
dominates -- see below). 

\subsection{Line emission from the extended bright-spot}

Balmer, He\,{\small I} and He\,{\small II} emission from 
the $(-V_x,-V_y)$ quadrant of Doppler maps seems to be a ubiquitous
feature of nova-like variables \cite{kaitchuck94}. This emission is 
not often associated with the bright spot, however, as it has a 
velocity which is usually too large in the $-V_y$ direction. In the 
case of SW Sex, it appears as if some of the emission does share the 
velocity of the bright spot, although most of it is still  
too large in the $-V_y$ direction. One possibility is that the latter
emission is actually due to post-shock material downstream (along the
disc edge) from the point where the gas stream strikes the disc, as has
recently been suggested for the short-period dwarf nova 
WZ Sge \cite{spruit97}. The resulting line emission, due to
recombination as the post-shock material cools down, would share the 
velocity of the outer edge of the disc. This then provides an 
explanation for the velocities in the extended bright-spot region of
SW Sex displayed in the Doppler maps of Fig.~\ref{fig:maps}, although 
a detailed radiative hydrodynamical calculation of the shock and
post-shock regions would be required to have confidence in such a
model. Another possibility is that the extended nature of the 
bright-spot emission in SW Sex is due to a bright-spot overflow, as 
suggested for V1315 Aql by \scite{hellier96}. We believe this is
unlikely, however, as one might then expect the emission to follow the
gas stream towards higher velocities on the Doppler map, unless
interaction with the disc is strong, in which case the emission would
be deflected in the $+V_y$ direction (towards the curve indicating the
Keplerian velocity of the disc along the gas stream in Fig.~\ref{fig:maps}). 

\subsection{The SW Sex phenomenon explained?}
\label{sec:conclusions}

The three-component model outlined above appears to have had some
success in explaining much of the behaviour exhibited by SW Sex. 
We are confident that the accretion disc and secondary star components
are correct, but find that the dominant bright-spot component fails
in one important area: its inconsistency with
the emission-line light curves. This can most readily be seen by 
inspecting the light curve of H$\gamma$ in Fig.~\ref{fig:lc}, where
the eclipse profile requires material to be eclipsed as early as
phase 0.8, but which is not as deeply eclipsed as the continuum,
exhibits a flat bottom and then comes out of eclipse very sharply around 
phase 0.05. The early ingress of the light curve 
(between phases $\sim0.8-0.9$) cannot be due to obscuration by the
secondary star, as the Roche lobe does not eclipse any part of the
disc during these phases. Instead, we attribute this to a raised
disc rim downstream from the bright-spot, 
as observed in the dwarf nova OY Car \cite{billington96}, which passes 
across the line of sight at this phase and which obscures the disc 
behind it\footnote{This might be used to explain another intriguing aspect 
of the SW Sex phenomenon: If the disc rim is raised to an angle above
the orbital plane which is larger than $90-i$, and if the material in
the raised rim is optically thick to the continuum emission from the
disc, then \scite{rutten97} has shown that the 
resulting eclipse maps will exhibit a flattening in the run of 
temperature with disc radius in the inner disc regions, as observed in
many nova-like variables \cite{rutten92b}.}. The rapid egress between 
phases 0.01 and 0.05, however, is difficult to explain with a dominant
bright-spot component as one would expect it to be obscured by the 
secondary at these phases (for any reasonable set of system
parameters). For the same reason, it is also difficult to explain the 
general reduction in line flux observed around phase 0.5.
It is still possible to account for the emission-line light curves 
with a dominant bright-spot component, however, by speculating that
there are regions of strong Balmer absorption in the disc whose changing
visibility during eclipse alters the shape of the light curve.
Alternatively, we might invoke some of the more exotic theories of the
past in which there is a source of Balmer emission from above the
orbital plane of the binary, although this would still have to share the 
velocity of the bright-spot. 

\section*{\sc Acknowledgments}

We are indebted to Ren\'{e} Rutten for the communication of his SW Sex 
eclipse data. TRM is supported by a PPARC Advanced Fellowship. The 
INT is operated on the island of La Palma by the RGO in the Spanish 
Observatorio del Roque de los Muchachos of the IAC.

\bibliographystyle{mnras}
\bibliography{abbrev,refs}

\end{document}